# An inter comparison of the upper critical fields ($H_{c2}$) of different superconductors - $YBa_2Cu_3O_7$, $MgB_2$, $NdFeAsO_{0.8}F_{0.2}$, $FeSe_{0.5}Te_{0.5}$ and $Nb_2PdS_5$

R. Sultana[1,#], P. Rani[1,2], A. K. Hafiz[2], Reena Goyal[1,#], and V. P. S. Awana[1,*]

[1]National Physical Laboratory (CSIR), Dr. K. S. Krishnan Road, New Delhi-110012, India

[2] Department of Physics, Jamia Millia Islamia, New Delhi-110025, India

**Abstract**

Here we report the comparison of the upper critical fields of different superconductors being calculated by two different theories i.e., Werthamer Helfand Hohenburg (WHH) and Ginzberg Landau (GL). All the samples are synthesized through previously known solid state reaction route. Phase purity is determined from the Rietveld refinement of powder X-Ray diffraction (XRD) data. High field (up to 14Tesla) magneto transport $\rho(T)H$ of different superconductors is studied to estimate their upper critical field ($H_{c2}$). The present inter comparison covers from Cuprates ($YBa_2Cu_3O_7$) - Borides ($MgB_2$) - Fe pnictides ($NdFeAsO_{0.8}F_{0.2}$) and chalcogenides ($FeSe_{0.5}Te_{0.5}$) to robust $Nb_2PdS_5$. The upper critical fields [$H_{c2}(T)$] at zero temperature are calculated by extrapolating the data using GL and WHH equations.

**Introduction**

No doubt the discovery of superconductivity [1] induced great interest in the field of condensed matter physics. A large number of superconducting compounds with unique properties have been discovered so far. The distinctive property of superconductors, expelling the external magnetic fields from interior and thus prevailing quantum oscillations do provide extraordinary measurement sensitivity and often the new physics [2]. As far as superconductivity is concerned, the upper critical field ($Hc_2$) plays an important role. The $Hc_2$ is an intrinsic parameter for any superconductor, determining as to how long the superconductivity remains under magnetic field, i.e., survival of cooper pairs (coherence length) and pairing potential strength [2-4]. For determination of $Hc_2$ the superconductor is subjected to high magnetic fields and its superconducting critical temperature ($T_c$) is measured at various fields. It is usually expressed in Tesla (T) and calculated by two different theories namely: Ginzburg Landau (GL) and Werthamer Helfand Hohenberg (WHH) theory. According to GL theory, superconductivity can be thought of as a phase transition involving a complex order parameter that goes to zero at $T_c$ and is an alternative to the London theory. The WHH predicts the upper critical field ($H_{c2}$) at 0K from decrement of $T_c$ under magnetic field and the slope of $H_{c2}$ at $T_c$. The $H_{c2}$ (T) is obtained using 10%, 50% and 90% criteria of the normal state resistivity value.

As far as the history of superconducting materials is concerned, the superconductivity research got a big boost in year 1986 after the discovery of high-$T_c$ superconductors including layered perovskites (LaBa,Sr)$_2$CuO$_4$ with T$_c$ of about 40K and YBa$_2$Cu$_3$O$_7$ with T$_c$ of 91K [5,6]. This further led to the discovery of new high-T$_c$ oxide superconductors e.g., the Bi-Sr-Ca-Cu-O system exhibiting a T$_c$ of above 105K [8] and an upper critical field (H$_{c2}$), exceeding 100Tesla. Similar structure Tl and Hg based cuprates were also discovered soon after, with their T$_c$ as high as up to 130K [8]. Numbers of investigations have been done on upper critical field of various cuprate High T$_c$ superconductors in the past [9-16].

In case of MgB$_2$, the upper critical field [H$_{c2}$(0)] of as high as 60Tesla has been reported in literature [17-20]. MgB$_2$ is a known type II superconductor with two superconducting gaps [18-22]. The discovery of superconductivity in Fe based (FeSCs) pnictides showed that FeSCs do exhibit extremely large upper critical fields [23,24]. For example for NdFeAsO$_{0.8}$F$_{0.2}$ superconductor, the H$_{c2}$(0) is reported to be above at least 100Tesla. As far as the robustness of superconductivity under magnetic field is concerned the Fe chalcogenides (FeSe) and the recently discovered Nb$_2$PdS$_5$ lead the pack with their upper critical fields at 0K i.e., H$_{c2}$(0) lying outside the Pauli paramagnetic limit [25,26]. The Pauli Paramagnetic limit is defined as $\mu_0H_p=1.84T_c$ [27]. This means the upper critical field of a superconductor at 0K i.e. H$_{c2}$(0) must be within 1.84T$_c$, where T$_c$ is the critical temperature of superconductor [4,27]. After giving short introduction to some of the exotic superconducting families starting from Cuprates to Borides and FeSCs including robust Nb$_2$PdS$_5$, it was thought appropriate to inter compare their upper critical fields at once place. This letter reports the inter comparison of the upper critical fields of various interesting superconducting families being synthesized in same laboratory at one place. Specifically, the high field (up to 14Tesla) transport measurements ρ(T)H are done in superconducting regime down to 2K for YBa$_2$Cu$_3$O$_7$, MgB$_2$, NdFeAsO$_{0.8}$F$_{0.2}$, FeSe$_{0.5}$Te$_{0.5}$ and Nb$_2$PdS$_5$ superconductors and their upper critical fields are estimated and compared.

**Experimental**

All the samples were synthesized through solid state reaction route at various temperatures as per the commonly reported literature. High purity reactants are accurately weighed in stoichiometric proportions and grounded properly to obtain homogeneously mixed powders with the help of mortar and pestle. The above step was performed in presence of high purity Argon atmosphere in glove box. The obtained, mixed powder was pelletized, vacuum-sealed (10$^{-4}$Torr) in a quartz tube and heated to desired temperatures. The furnace was then allowed to cool down slowly to room temperature. Sintered samples obtained were in the form of black powder and of dense form which were good enough for further transport measurements. In case of YBa$_2$Cu$_3$O$_7$ the heat treatment programme is carried out in open i.e., without quartz encapsulation. The structural characterization of the synthesized samples was done through room temperature X-ray diffraction using Rigaku X-ray diffractometer with Cu K$_\alpha$ radiation ( λ=1.5418 Å ) for checking phase purity (structural analysis). The magnetic and resistivity measurements under magnetic field were carried out by a conventional four-probe method on a quantum design Physical Property Measurement System (PPMS) with fields up to 14Tesla.

**Results and discussion**

Figure 1 shows the Rietveld fitted room temperature XRD pattern of YBa$_2$Cu$_3$O$_7$, FeSe$_{0.5}$Te$_{0.5}$, Nb$_2$PdS$_5$, MgB$_2$ and NdFeAsO$_{0.8}$F$_{0.2}$ being labelled as (a), (b), (c), (d) and (e), respectively. Purity of all the samples can be clearly seen from their respective refined XRD patterns. Except MgB$_2$, where small amount of MgO is seen as an impurity, all other samples are mostly crystallized in single phase having different space group. The YBa$_2$Cu$_3$O$_7$ sample is crystallized in orthorhombic structure within Pmmm space group. Both the samples (b)

FeSe$_{0.5}$ Te$_{0.5}$ and (e) NdFeAsO$_{0.8}$F$_{0.2}$ are crystallized in tetragonal structure within P4/nmm space group. Sample (c) Nb$_2$PdS$_5$ is crystallized in mono-clinic structure with *C*2/*m* (#12) space group. Sample (d) MgB$_2$ is fitted in P6/mmm space group. The XRD pattern of MgB$_2$ sample showed small trace of MgO at 2theta≈63˚.The lattice parameters of all the samples are shown in table 1. It is clear that the studied polycrystalline superconducting samples from various families are mostly single phase in nature.

Figure 2 (a-e) shows the temperature dependence of resistivity of all the studied samples, i.e., YBa$_2$Cu$_3$O$_7$, FeSe$_{0.5}$Te$_{0.5}$, Nb$_2$PdS$_5$, MgB$_2$ and NdFeAsO$_{0.8}$F$_{0.2}$ respectively under high magnetic fields of up to 14Tesla. The field is applied perpendicular to the direction of the current flow. In the absence of magnetic field the YBa$_2$Cu$_3$O$_7$ sample exhibited sharp superconducting transition in a single step as compared to the one being under applied magnetic field. The onset transition value at zero magnetic field i.e., T$_c^{onset}$ is around 91K. When magnetic field is applied, the onset transition temperature remains nearly unchanged while, T$_c$ ($\rho$=0) which is around 88K, decreases to a lower temperatures. Under applied magnetic field of 13Tesla, the T$_c$ ($\rho$=0) of YBa$_2$Cu$_3$O$_7$ decreases to below 45K. Further, the decrease of T$_c$ ($\rho$=0) is not in single step but looks like a two step transition due to the weak links which we have been discussed in detail our earlier report [28].

Part (b) of Figure 2 shows the field dependence of resistivity of FeSe$_{0.5}$Te$_{0.5}$ sample under applied fields of up to 9Tesla. In absence of field the superconducting transition temperature is around 15.1K, where as the offset value of transition temperature is around 12.4K. Interestingly, the behaviour of this sample under magnetic field is not similar to the previous sample i.e., YBa$_2$Cu$_3$O$_7$. Both the onset and offset temperatures are field dependent and decrease to lower temperatures with increase in magnetic field. Onset is slightly less affected than the offset. Under 9Tesla magnetic field the onset value decreases to 13.27K and the offset drops to 8.2K. Unlike the case of YBa$_2$Cu$_3$O$_7$, the transition in FeSe$_{0.5}$Te$_{0.5}$ under high magnetic field up to 9Tesla is single step. This is due to strong granular coupling within the grains of this system.

Part (c) and (d) show the resistivity behaviours of Nb$_2$PdS$_5$ and MgB$_2$ respectively. The general trend is similar to that as in case of FeSe$_{0.5}$Te$_{0.5}$. Both the onset and offset temperature shifts almost parallel towards the lower temperatures with increase in magnetic field. The transition in case of MgB$_2$ sample is sharp as compared to other samples. In case of MgB$_2$ sample the onset temperature at zero field is around 37.7K and the offset is at 36.3K. The transition width ($\Delta$T) is 1.4K. Under 8Tesla magnetic field onset shifts to 23.5K and the offset temperature decreases to 15.3K with an increased transition width of 8.2K. In case of Nb$_2$PdS$_5$, under the absence of magnetic field the onset temperature is seen at around 6.69K and offset is at 5.84K with transition width of 0.8K. Under magnetic field of 14Tesla, the onset decreases to 4.04K and offset shifts to 2.38K. Therefore, the calculated transition width comes out to be 1.7K.

In case of sample NdFeAsO$_{0.8}$F$_{0.2}$ (Fig. 2e), the onset transition T$_c^{onset}$ is seen at around 51.8K and the offset temperature T$_c^{offset}$ is at around 46.8K. Under applied magnetic field of 14Tesla, the T$_c^{offset}$ shifts towards the lower temperature and decreases to below 30.3K, with an onset value of still around 50K. Nearly unchanged value of T$_c^{offset}$ under magnetic field for NdFeAsO$_{0.8}$F$_{0.2}$ (Fig. 2e) reminds the similar situation being observed for YBa$_2$Cu$_3$O$_7$ (Fig. 2a). However, there is a striking difference, i.e., the two step transition, which is seen clearly in case of YBa$_2$Cu$_3$O$_7$ sample (Fig. 2a) is not that transparent in case of NdFeAsO$_{0.8}$F$_{0.2}$ (Fig. 2e).

Summarising, the magneto-transport results in superconducting regime for different class of superconductors, it is clear that their response to applied magnetic field is not exactly similar to each other. Not only the relative decrease in superconducting transition, but the broadening of transition under magnetic field is strikingly different.

To know the upper critical field of various studied superconductors, here we have calculated the upper critical field of different superconductors by two different models i.e., WHH and GL. Figure 3(a-e) shows the upper critical field [$H_{c2}(T)$] for $YBa_2Cu_3O_7$, $FeSe_{0.5}Te_{0.5}$, $Nb_2PdS_5$, $MgB_2$ and $NdFeAsO_{0.8}F_{0.2}$ at zero temperature as calculated by extrapolating the data using Ginzburg-Landau (GL) equation. The $H_{c2}(T)$ is obtained using 10%, 50% and 90% criteria of $\rho_N$ (normal state resistivity) i.e., where the resistivity becomes 10%, 50% and 90% of its normal state value. The Ginzburg-Landau equation is given as,

$$H_{c2}(T) = H_{c2}(0) * \left[\frac{(1-t^2)}{(1+t^2)}\right]$$

Where $t = T/T_c$ is the reduced temperature [24] and $H_{c2}(0)$ is the upper critical field at zero temperature. As shown in the figure 3 (a) - (e) the $NdFeAsO_{0.8}F_{0.2}$ has a large $H_{c2}$ value compared to the other three samples except $YBa_2Cu_3O_7$, which possesses even higher value. Rest of the studied superconducting compounds upper critical field values are less than 100Tesla. $Nb_2PdS_5$ is a known robust compound, still it has small upper critical field, because of its lower $T_c$ of around 6K only. WHH theory gave a solution for linear Gorkov equations for $H_{c2}$ for type-II superconductor [29]. In order to estimate the $H_{c2}(T)$ value we use the simplified WHH equation as given below

$$H_{c2}(0) = -0.693 \left(\frac{dH_{c2}}{dT}\right) T_c$$

where, $H_{c2}(0)$ is the upper critical field at zero temperature, $\left(\frac{dH_{c2}}{dT}\right)$ denotes the slope of $H_{c2}(T)$ near $T_c$ [29-31]. Here, we considered the onset transition temperature as the upper critical field $T_c(H_{c2})$. Similar to the GL theory, we calculate the upper critical field value by taking a criterion of 10%, 50% and 90% of $\rho_N$ (normal state resistivity) i.e., where the resistivity becomes 10%, 50% and 90% of its normal state value. It is seen that the slope of $H_{c2}(T)$ near $T_c$, i.e. $dH_{c2}/dT$ for $YBa_2Cu_3O_7$ are about −0.322 Tesla/K, -0.845Tesla/K and -6.254Tesla/K for 10%, 50% and 90% respectively. Similarly for $FeSe_{0.5}Te_{0.5}$ the $dH_{c2}/dT$ values at 10%, 50% and 90 % are -2.519Tesla/K, -3.453Tesla/K and -4.411Tesla/K respectively. For $Nb_2PdS_5$ the values obtained are -3.405Tesla/K,-3.688Tesla/K and -4.426Tesla/K. For $MgB_2$ the values are -0.4239Tesla/K, -0.4598Tesla/K and -0.5233Tesla/K. Finally the values calculated for $NdFeAsO_{0.8}F_{0.2}$ are -0.573Tesla/K, -2.734Tesla/K and -6.415Tesla/K. The calculated values of $H_{c2}$ for various superconductors i.e., $YBa_2Cu_3O_7$, $MgB_2$, $NdFeAsO_{0.8}F_{0.2}$, $FeSe_{0.5}Te_{0.5}$ and $Nb_2PdS_5$ with 10%, 50% and 90% criterion of $\rho_N$ are given in Table 2. It is found that the $H_{c2}$ calculated by WHH model are less than the corresponding values calculated by GL model. The earlier reported theoretical model [4] suggested upper critical field limit for hard superconductor as $H_0 = 1.84T_c$ in Tesla, i.e., so called Pauli paramagnetic limit to break the cooper pairs. Interestingly, both the $FeSe_{0.5}Te_{0.5}$ and $Nb_2PdS_5$ superconductors are clearly outside the $H_0 = 1.84T_c$ limit even with the 10% criterion of $\rho_N$. On the other hand the upper critical field of Cuprate ($YBa_2Cu_3O_7$) and the Fe-Pnictide ($NdFeAsO_{0.8}F_{0.2}$) also crosses the $H_0 = 1.84T_c$ limit if one considers the 90% criterion of $\rho_N$. This is so because the onset temperature of superconductivity in $YBa_2Cu_3O_7$ and $NdFeAsO_{0.8}F_{0.2}$ is hardly affected by the applied magnetic field. The case of $MgB_2$ is seen well within the $H_0 = 1.84T_c$ limit with $\rho_N$ criterion of 10%, 50% or 90%.

Seemingly, though $MgB_2$ seems to follow the conventional rules of superconductivity, the $YBa_2Cu_3O_7$, $FeSe_{0.5}Te_{0.5}$, $Nb_2PdS_5$, and $NdFeAsO_{0.8}F_{0.2}$ are different and hence considered to be the exotic ones.

**Acknowledgements**

Authors would like to thank their Director Professor D.K. Aswal for his keen interest and encouragement in the present work. R. Sultana thanks CSIR, India for providing research fellowship. This work was supported by DAE-SRC outstanding investigator award scheme to work on search for new superconductors.

**Table captions**

**Table 1:** Rietveld refined lattice parameters with corresponding transition temperature.

**Table 2:** Calculated Upper critical field ($H_{c2}$) of different superconductors

**Figure caption**

**Figure 1:** Rietveld fitted room temperature X-ray diffraction patterns for representative samples of (a) $YBa_2Cu_3O_7$, (b) $FeSe_{0.5}Te_{0.5}$, (c) $Nb_2PdS_5$, (d) $MgB_2$ and (e) $NdFeAsO_{0.8}F_{0.2}$.

**Figure 2:** Resistivity behaviour with temperature variation $\rho(T)$ of representative samples of (a) $YBa_2Cu_3O_7$, (b) $FeSe_{0.5}Te_{0.5}$, (c) $Nb_2PdS_5$, (d) $MgB_2$ and (e) $NdFeAsO_{0.8}F_{0.2}$.

**Figure 3:** The Ginzburg Landau (GL) fitted variation of upper critical field ($H_{c2}$) with temperature for (a) $YBa_2Cu_3O_7$, (b) $FeSe_{0.5}Te_{0.5}$, (c) $Nb_2PdS_5$, (d) $MgB_2$ and (e) $NdFeAsO_{0.8}F_{0.2}$ samples respectively at 90%, 50% and 10% criteria of the normal state resistivity value.

**Table 1:** Rietveld refined lattice parameters with corresponding transition temperature.

| Compound | a(Å) | b(Å) | c(Å) | Vol.(Å$^3$) | $\chi^2$ | $T_c$ (K) |
|---|---|---|---|---|---|---|
| YBa$_2$Cu$_3$O$_7$ | 3.829(3) | 3.887(4) | 11.675(5) | 173.838 | 2.93 | 90.7 |
| FeSe$_{0.5}$Te$_{0.5}$ | 3.798(5) | 3.798(5) | 5.992(3) | 86.462 | 4.87 | 14 |
| Nb$_2$PdS$_5$ | 12.21(3) | 3.27(2) | 15.23(7) | 569.923 | 8.1 | 5.7 |
| MgB2 | 3.063(2) | 3.063(5) | 3.503(6) | 28.485 | 6.4 | 37.8 |
| NdFeAsO$_{0.8}$F$_{0.2}$ | 3.971(2) | 3.971(2) | 8.571(3) | 135.17 | 2.54 | 51.8 |

**Table 2:** Calculated Upper critical field ($H_{c2}$) of different superconductors

| Sample | | $H_{c2}$ (Tesla) | | |
|---|---|---|---|---|
| | Theory used | 10% $\rho_N$ | 50% $\rho_N$ | 90% $\rho_N$ |
| YBa$_2$Cu$_3$O$_7$ | WHH | 19.698 | 52.496 | 393.91 |
| | GL | 24.62 | 70.4493 | 563.091 |
| FeSe$_{0.5}$Te$_{0.5}$ | WHH | 22.912 | 33.272 | 45.205 |
| | GL | 28.47 | 43.051 | 60.012 |
| Nb$_2$PdS$_5$ | WHH | 14.055 | 15.623 | 19.659 |
| | GL | 16.7 | 18.822 | 24.475 |
| MgB$_2$ | WHH | 10.824 | 11.824 | 13.545 |
| | GL | 12.564 | 13.824 | 16.1 |
| NdFeAsO$_{0.8}$F$_{0.2}$ | WHH | 55.538 | 92.964 | 225.542 |
| | GL | 71.459 | 127.45 | 318.195 |

**Figure 1**

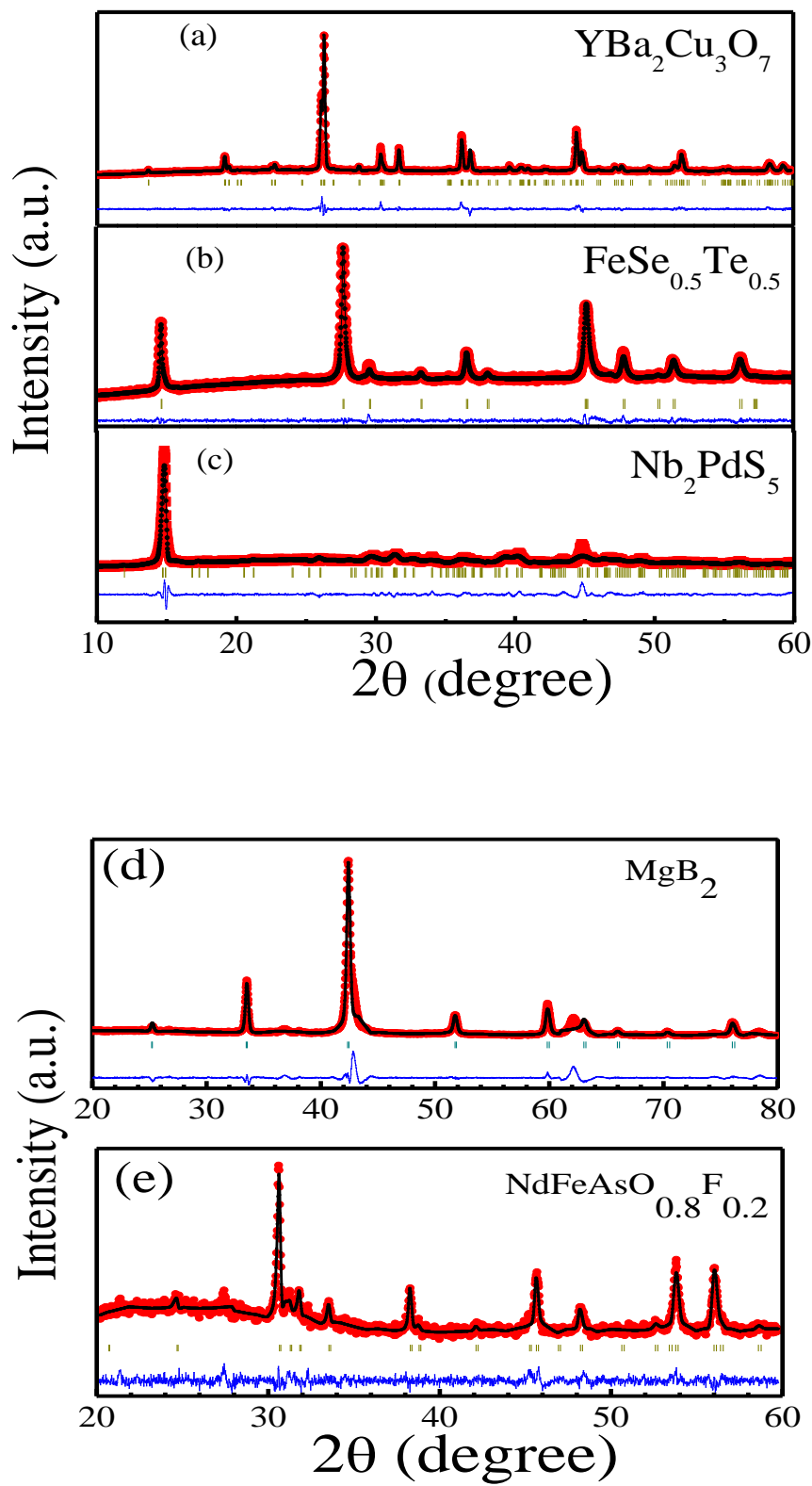

**Figure 2**

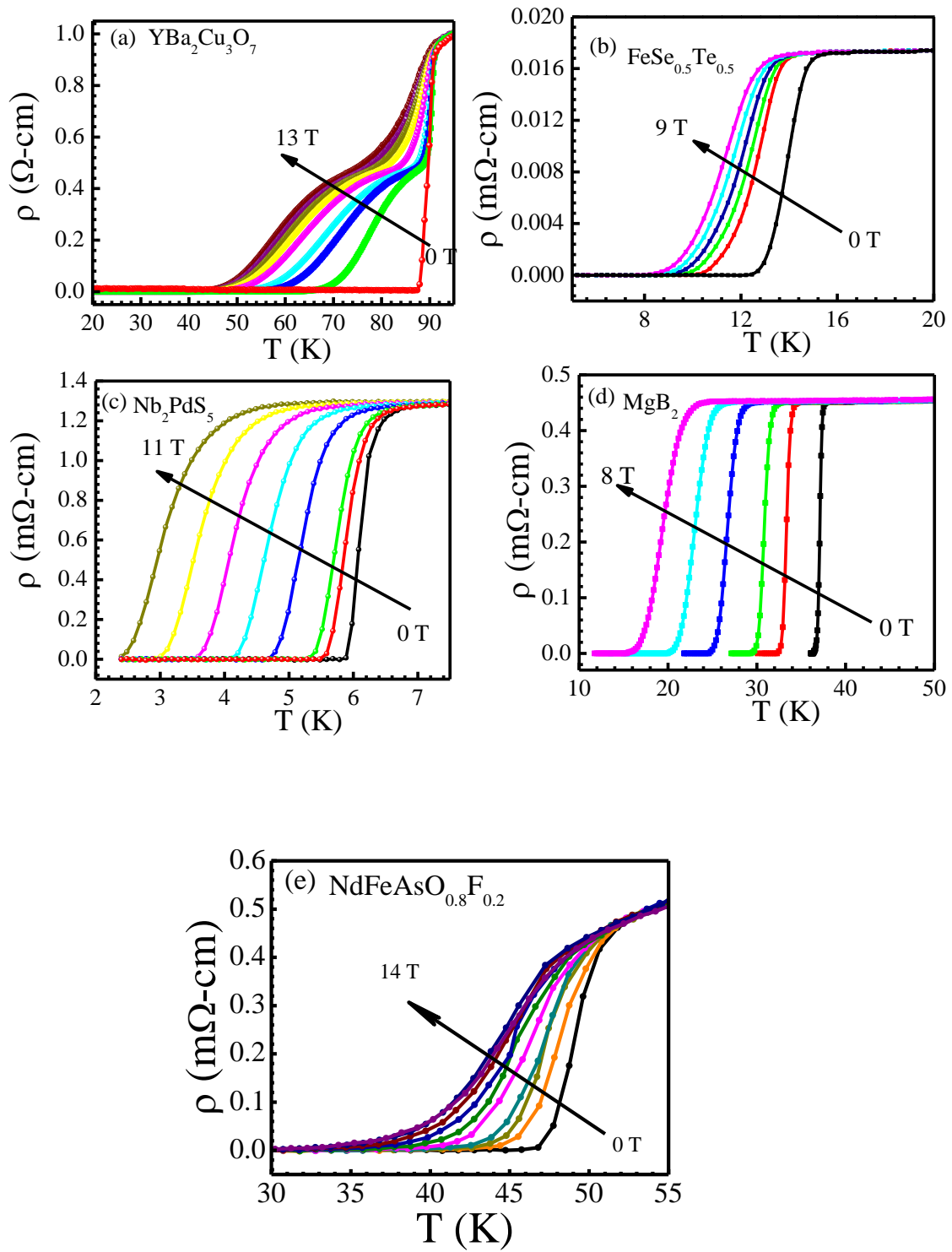

**Figure 3**

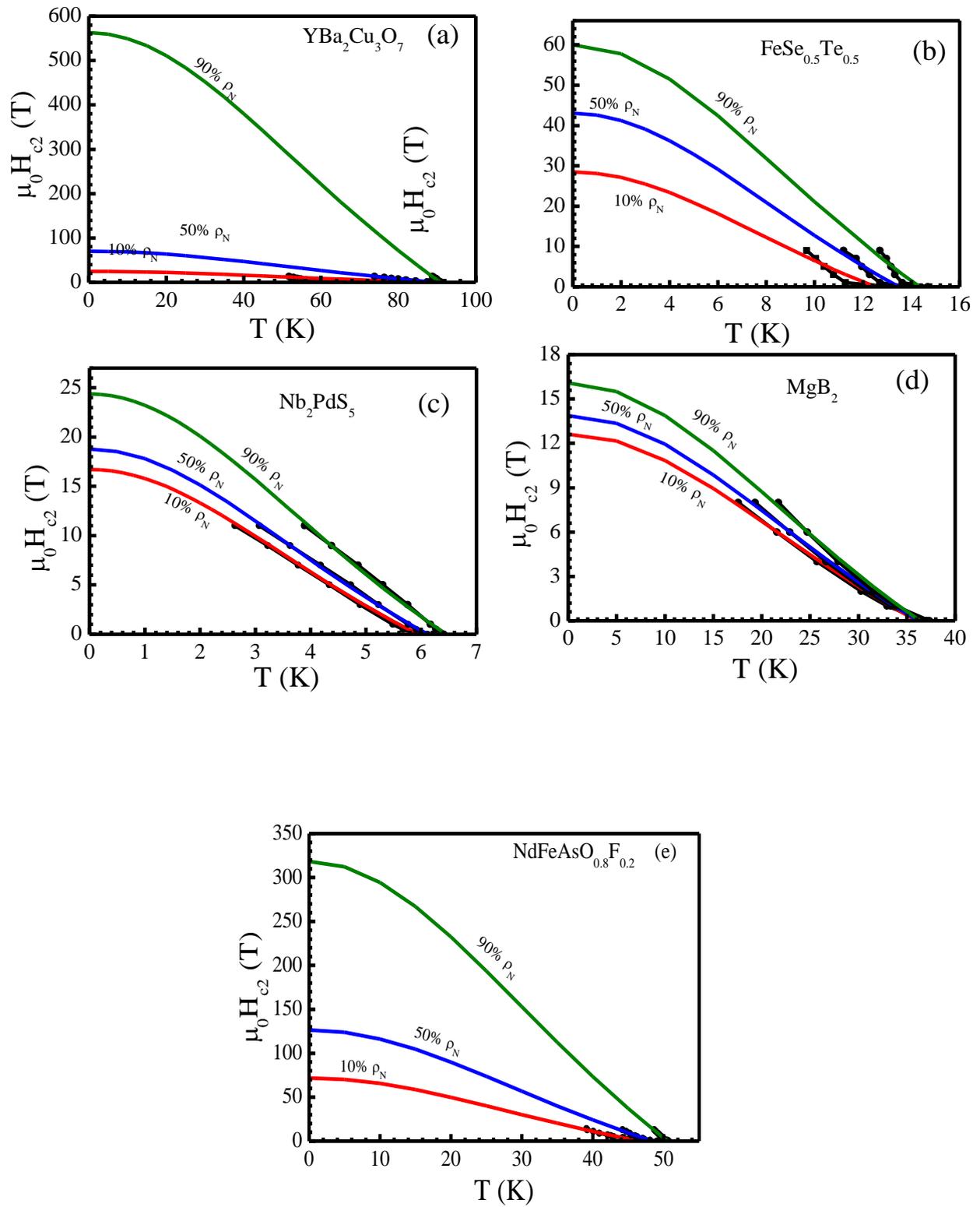